# Interaction between interface and massive states in multivalley topological heterostructures

G. Krizman[1], B. A. Assaf[2], M. Orlita[3,4], G. Bauer[1], G. Springholz[1], R. Ferreira[5], L. A. de Vaulchier[5], Y. Guldner[5]

[1] *Institut für Halbleiter und Festkörperphysik, Johannes Kepler Universität, Altenbergerstraße, 69, 4040 Linz, Austria*

[2] *Department of Physics, University of Notre Dame, Notre Dame, IN 46556, USA*

[3] *LNCMI, CNRS-UGA-UPS-INSA-EMFL, 25, avenue des Martyrs, F-38042 Grenoble, France*

[4] *Institute of Physics, Charles University, CZ-12116 Prague, Czech Republic*

[5] *Laboratoire de Physique de l'Ecole Normale Supérieure, ENS, Université PSL, CNRS, Sorbonne Université, 24 rue Lhomond 75005 Paris, France*

**ABSTRACT**

**Topological interface states in multivalley systems are studied to unravel their valley sensitivity. For this purpose, multivalley IV-VI topological crystalline insulator (TCI) heterostructures are explored using magneto-optical Landau level spectroscopy up to 34 teslas. We characterize the topological interface states emerging from the distinct $L$-valleys in $Pb_{1-x}Sn_xSe$ multi quantum wells grown along the [111] direction. It is shown that the shape of the 2D Fermi surfaces of topological interface states residing at the TCI/trivial insulator interfaces are strongly affected by the valley anisotropy of topologically trivial $Pb_{1-y}Eu_ySe$ barriers. This phenomenon is shown to be due to the deep penetration of the topological interface states into the barriers. For the valleys tilted with respect to the confinement direction, a significant interaction between topological states and the conventional massive quantum well states is observed, evidenced by the resulting large anti-crossings between Landau levels. These are theoretically well-described by a $k.p$ model that takes into account tilt and anisotropy of the valleys in two dimensions. Therefore, our work provides a precise characterization of the topological interface state valley splitting, as well as an accurate determination of the anisotropy of their Dirac cone dispersion.**

## I. INTRODUCTION

Multivalley semiconductors offer additional degrees-of-freedom for tuning of the electronic properties for novel "valleytronics" device applications because these properties significantly depend on the structure of the valleys as well as their interaction [1–5]. It is, however, a major challenge to control and determine the valley splitting and valley anisotropy in semiconductors. For instance, qubit manipulation in multivalley systems like Si, Si/Si$_{1-x}$Ge$_x$ or Ge/Ge$_{1-x}$Si$_x$ heterostructures, which is based on spin manipulation, requires a fine-tuned valley splitting to ensure Pauli blockade and avoid decoherence [6–8]. An optimized valley splitting has recently allowed to obtain qubits at an enhanced operating temperature [9,10]. The additional valley degree-of-freedom can also be used to control the electric current and constitutes the basis of valleytronics device applications such as valley-based classical bits [11,12] and/or qubits [13,14]. From a fundamental point of view, the valley anisotropy is also at the origin of spontaneous crystal symmetry breakings, leading to a novel nematic phases, or the quantum Hall ferroelectric state [15–18]. The lead salt semiconductors are a particularly interesting template for valleytronics not only because of their multivalley band structure [19,20] but also because through alloying with tin they can be converted into topological crystalline insulator (TCI) featuring massless and spin polarized Dirac electrons with the spin locked to the momentum [21–27]. Therefore, they can be envisioned for topological valleytronics in which information is stored, e.g., by the different chirality of electrons in different valleys.

In the lead salt compounds, the band extrema are located at the four $L$-points of their Brillouin zone (BZ), as illustrated in Fig. 1(a) [19,28], where in each valley, the band structure consists of two mirror bands $L_6^+$ and $L_6^-$ with opposite symmetry (see Fig. 1(b)) [19,29,30]. In pseudo binary Pb$_{1-x}$Sn$_x$Se and Pb$_{1-x}$Sn$_x$Te alloys, this symmetry can be inverted when the Sn content exceeds a certain critical value [20,22–24,31–33]. This leads to the emergence of the TCI phase, whose hallmark is the existence of Dirac (interface or surface) states protected by crystalline symmetry at the (001) and (111) crystal faces [34,35]. Due to the multivalley band structure, for the (111) oriented surface three Dirac cones appear at the $\bar{M}$-points of the 2D BZ and one is located at the center $\bar{\Gamma}$, as depicted in Fig. 1(a). The $\bar{M}$ and $\bar{\Gamma}$-points can be regarded as the projection of the four $L$-points in the 3D BZ to the 2D BZ of the (111) surface, corresponding to the oblique and longitudinal valleys, respectively. Extensive studies have been performed to quantify the valley anisotropy in bulk lead salt materials [33,36–39] and at free surfaces of Pb$_{1-x}$Sn$_x$Se TCIs using surface-sensitive techniques [24,27,40]. Here, we focus on the topological states at buried interfaces of topological multilayer structures probed by magnetooptical spectroscopy to accurately characterize both interface and bulk states, thereby revealing the significant impact of the bulk on the valley splitting and anisotropy in heterostructure systems.

The electronic properties of the lead salt compounds are well-described by a 4-band $\boldsymbol{k}.\boldsymbol{p}$ Hamiltonian, which is similar to a massive anisotropic 3D Dirac Hamiltonian [33,41]. The constant energy surfaces are ellipsoids with an anisotropy factor $K$ defined by the ratio between the main and minor axes of the ellipsoids (see Fig. 1(a)) [39,42]. Although $K$ being identical for the four valleys, their different locations within the BZ lead to different orientations of their major axes. One ellipsoid has its long main axis aligned along the [111] direction and the three others are tilted by an angle $\theta = 70.5$ ° with respect to [111]. We will refer to these as longitudinal and oblique valleys, respectively, as drawn in Fig. 1(a).

Here, we focus on the multivalley features of multi quantum well (MQW) heterostructures composed of topologically non-trivial Pb$_{0.75}$Sn$_{0.25}$Se quantum wells separated by topologically trivial barriers of Pb$_{1-y}$Eu$_y$Se, as shown schematically in Fig. 1(b) [22,24,33,43]. In addition to the usual quantum confined states localized within the QW (orange states in Fig. 1(b,c)), the switching of the topological character at each interface gives rise to topological interface states (TIS) localized at the hetero-interfaces, shown in blue in Fig. 1(b,c) and already observed in our previous work [44]. The relatively small thicknesses of the wells yield hybridization between top and bottom TIS of each QW that opens a gap $2\delta$ in the Dirac cone, between the electron-like and the hole-like TIS, here denoted as TIS and TIS' respectively [44–46]. Surprisingly, using magneto-optical Landau level spectroscopy, we show that TIS and TIS' are very sensitive to the valley anisotropy of the bulk barrier material, due to their remarkable property of being localized mainly at the interface, contrary to trivial QW states found in ordinary semiconductor heterostructures. Even though the quantum well material $Pb_{0.75}Sn_{0.25}Se$ has rather isotropic valleys, the valley anisotropy of the barrier material $Pb_{1-y}Eu_ySe$ is sufficient to strongly impact the dispersion of the TIS. As a result, the interaction between the TIS Landau levels and those of trivial QW states is allowed and generates an observable signature in Landau level spectroscopy. Lastly, we show that this impact is unique to the $\bar{M}$-points of the 2D Brillouin zone and does not occur at the $\bar{\Gamma}$-point.

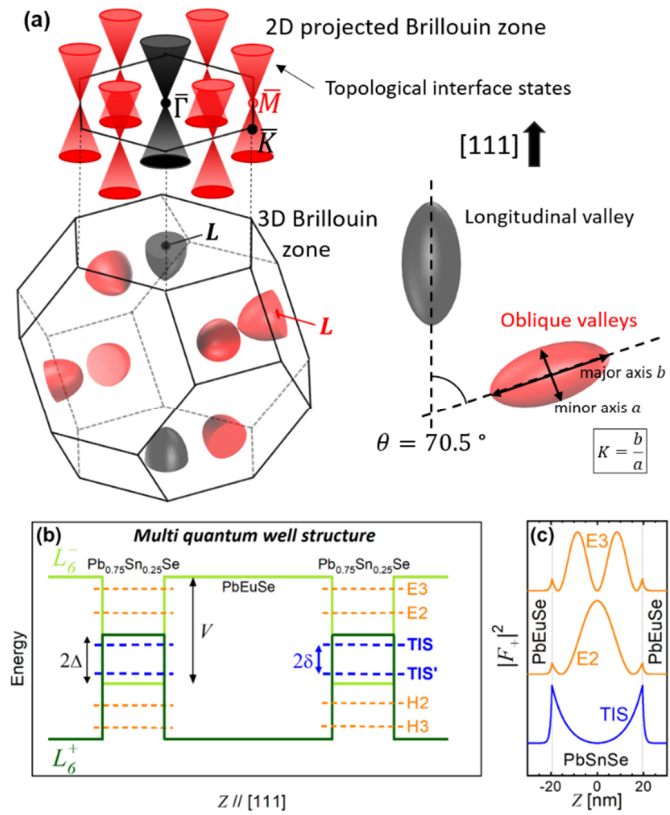

**Figure 1.** (a) 3D Brillouin zone of cubic lead salt crystals and its 2D projection onto the (111) surface. Band minima in 3D are located at the four different $L$-points: 3 in red and 1 in black. The surfaces of constant energy of these band minima are plotted for the two types of valleys with respect to [111]: the longitudinal valley depicted in black and the three equivalent oblique valleys in red, tilted by $\theta = 70.5\,°$ with respect to the surface normal [111]. They both have an anisotropic factor $K$, defined as the ratio between the major and minor axis of the ellipsoids. The projection onto the (111) surface corresponds to the location of the TIS Dirac cones. Three at the $\bar{M}$-points in red stem from the band inversion in the oblique valleys and one at the $\bar{\Gamma}$-point in black originates from the longitudinal valley. Note that these cones can be gapped with a hole part TIS' and an electron part TIS. (b) Schematic band alignment of $L_6^\pm$ bands between $Pb_{0.75}Sn_{0.25}Se$ (well) and $Pb_{1-y}Eu_ySe$ (barrier) along $Z$ parallel to the [111] growth direction. The confined state energy levels at $k_\parallel = 0$ are represented by dashed lines in blue for TIS and TIS' and in orange for massive states, labeled H2, H3, E2, E3, … (c) Square of the $F_+$-component of the envelope function spinor for TIS in blue and the two first massive confined states E2 and E3 in orange calculated using the parameters for the sample MQW-39 - see Table I. The curves are vertically shifted for clarity.

The paper is divided as follows. In Sec. II, the experimental results are introduced. We develop in Sec. III a complete theory for the confinement effect on a 3D anisotropic Dirac model, considering multivalley band structure, which applies for lead salts in particular. We show that the anisotropy and valley tilt of the parent bulk materials allow for a finite interaction between TIS and massive confined quantum well states (indicated in blue and orange states in Fig. 1(b,c)), lifting valley degeneracy at some particular magnetic fields or in-plane momenta. The analysis of the experimental results using the theory is presented in Sec. IV. We experimentally demonstrate that this anisotropy induces interactions between these states, observable as a set of repeated anti-crossings that occur between the Landau levels of topological and trivial states. Our spectroscopy also allows to quantify the strength of the interaction potential between different levels and to accurately deduce the in-plane anisotropy of the TIS Dirac cone dispersion [20,47].

## II. PESENTATION OF THE EXPERIMENTS

The investigated samples were grown by molecular beam epitaxy on freshly cleaved (111) $BaF_2$ substrates following the procedures detailed in our previous works [44,48]. Two MQWs were prepared with ~15 periods of $Pb_{0.75}Sn_{0.25}Se$ (well material) and $Pb_{1-y}Eu_ySe$ (barrier material) to which we refer as MQW-39 and MQW-25 according to their well thicknesses (see Table I). Note that the barrier material composition is $y = 0.05$ and $y = 0.1$ for MQW-25 and MQW-39 respectively. As a reference sample, a 2-$\mu$m thick film of $Pb_{0.75}Sn_{0.25}Se$ was grown. The Sn content $x = 0.25 \pm 0.01$ in the three samples insures a topological phase for temperature below ~100 K [33,49,50]. The ratio between the barrier and the well thicknesses is fixed as $5:1$ to suppress any coupling of confined states between adjacent quantum wells. To obtain a low carrier concentration in the three samples, Bi doping was employed during growth to compensate the native $p$-type character of $Pb_{0.75}Sn_{0.25}Se$ due to Se vacancies. Carrier densities as low as few $10^{17}$ cm$^{-3}$ and mobilities above $10^4$ cm$^2$/Vs are obtained in these samples [44,51].

**Table I.** Structure parameters of the $Pb_{0.75}Sn_{0.25}Se$ bulk reference sample and the two $Pb_{0.75}Sn_{0.25}Se$ TCI MQWs with $Pb_{1-y}Eu_ySe$ barriers used in the present investigations.

| Parameter | Bulk | MQW-39 | MQW-25 |
|---|---|---|---|
| $d_{PbSnSe}$ [nm] | 2000 | 39±2 | 25±1 |
| $x_{Sn}$ | 0.25±0.01 | 0.25±0.01 | 0.25±0.01 |
| $d_{PbEuSe}$ [nm] | N.A | 200 | 120 |
| $y_{Eu}$ | N.A | 0.10±0.01 | 0.05±0.01 |
| Period number | N.A | 15 | 16 |

Infrared magneto-spectroscopy is used to characterize the electronic properties of $Pb_{1-x}Sn_xSe$ MQWs and the TIS band structure. Experiments in magnetic fields up to 34 T are performed at the Laboratoire National des Champs Magnétiques Intenses in Grenoble. A setup as previously described in Ref. [52] is used to measure the infrared transmission of the samples with a magnetic field $B$//[111] (Faraday geometry). The transmitted radiation through the samples is collected in a same bath by a composite Si bolometer operating at 1.8 K, which is thus also the sample temperature.

Figure 2 compares the magneto-optical fan charts obtained on the bulk sample (Fig. 2(a)) to that of MQW-39 (Fig. 2(b)). The dots are obtained by pin-pointing the energy of the experimentally determined transmission minima and plotting these versus magnetic field and the solid lines in Fig.

2(a) represent the Landau level transitions calculated by solving the 4-band $\mathbf{k}\cdot\mathbf{p}$ Hamiltonian for bulk material, as described below. In the Faraday geometry, the interband transitions obey the selection rules $n \rightarrow n \pm 1$ with a Landau level spin flip [53]. The data are in perfect agreement to our model using the parameters listed in Table II. For the given composition, we obtain a negative bulk band gap $2\Delta = -52.5$ meV, in good agreement with previous studies [25,33,49,54]. Only one series of Landau levels with $v_\parallel = 4.50 \times 10^5$ m/s is required to explain our results, meaning that the longitudinal and oblique valleys are indistinguishable in the bulk material, i.e., the constant energy surfaces are approximately spheres ($K \sim 1$).

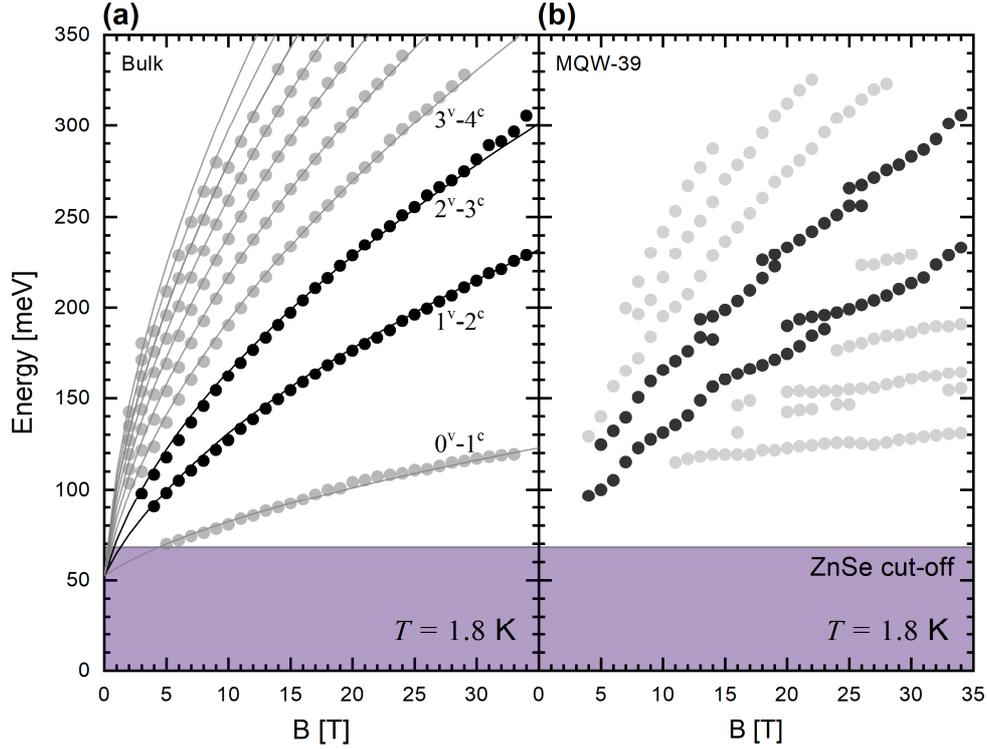

**Figure 2.** Magneto-optical transition fan charts for the bulk-like $Pb_{0.75}Sn_{0.25}Se$ film **(a)** and MQW-39 **(b)**. Dots represent the energy of transmission minima plotted versus applied magnetic field. For MQW-39, two series of transitions are highlighted to emphasize the difference between (a) and (b). In (a), the solid lines represent the calculated interband Landau level transitions for the thick film sample that perfectly fit to the experimental data, whereas for the MQW the dependence is highly non-monotonic. The interband transitions are labelled by the indices $n$ of the two involved Landau levels. The shaded area denotes the experimentally non-reachable range of photon energies below 60 meV due to energy cut-off of the ZnSe windows used in the experimental setup.

Comparing Fig. 2(a) and (b), a striking difference between the fan charts of the bulk and the MQW-39 samples is observed. Whereas the bulk sample displays a *high regularity* of the transmission minima versus magnetic field that evolves as $\sim \sqrt{B}$ as expected for bulk Dirac materials, in the MQW a pronounced *non-monotonic* dependence of the interband transition energies as a function of $B$ emerges. In the following, we will show that this behavior arises from an interaction between TIS and massive confined states in the MQW. This coupling is an inherent consequence of the topological and valley degenerate character of $Pb_{1-x}Sn_xSe$.

## III. THEORY

The electronic properties of lead salts including Pb$_{1-x}$Sn$_x$Se or Pb$_{1-y}$Eu$_y$Se can be well described by a 4-band $\bm{k}\cdot\bm{p}$ Hamiltonian, written near the $L$ points on the basis of the $L_6^+$ and $L_6^-$ conduction and valence bands [19,33,42]:

$$H = \begin{pmatrix} -\left[\Delta + \dfrac{p_\parallel^2}{2\widetilde{m}} + \dfrac{p_z^2}{2\widetilde{\mu}}\right]\mathbb{1} & v_\parallel \bm{p}_\parallel\cdot\bm{\sigma}_\parallel + v_z p_z \sigma_z \\ v_\parallel \bm{p}_\parallel\cdot\bm{\sigma}_\parallel + v_z p_z \sigma_z & \left[\Delta + \dfrac{p_\parallel^2}{2\widetilde{m}} + \dfrac{p_z^2}{2\widetilde{\mu}}\right]\mathbb{1} \end{pmatrix} \tag{1}$$

In this equation, $\mathbb{1}$ is the 2x2 identity matrix and $\bm{\sigma}$ the Pauli matrices. $2\Delta$ denotes the bulk band gap of the system, which is positive for Pb$_{1-y}$Eu$_y$Se and negative for Pb$_{0.75}$Sn$_{0.25}$Se at $T = 1.8$ K. $\widetilde{m}$ represents the contribution of all the remote bands to the in-plane effective mass. It exerts rather a small influence as these are at about 1 eV above and below $L_6^+$ and $L_6^-$ [30]. Note that the far-bands correction $\widetilde{\mu}$ to the longitudinal effective mass can be safely neglected as it has been demonstrated to be very small in the lead salt compounds [53]. In Eq. (1), the anisotropy is encoded by the in-plane and longitudinal Dirac velocities $v_\parallel$ and $v_z$ respectively, and thus, the anisotropy factor of bulk states is given by $K = v_\parallel/v_z$. The Hamiltonian is written for a $z$ axis parallel to the major axis of the ellipsoid.

### A. Bulk Landau levels

In order to interpret the magneto-optical data with the magnetic field parallel to [111], we calculate in the following the Landau levels. For the longitudinal valley, the $z$ axis is aligned with the [111] direction and therefore with $\bm{B}$. $p_x$ and $p_y$ are expressed in terms of the ladder operators following the Peierls substitution. The Hamiltonian at $k_z = 0$ is given by:

$$H = \begin{pmatrix} -\Delta - \hbar\widetilde{\omega}(n-1) & 0 & 0 & v_\parallel\sqrt{2e\hbar B n} \\ 0 & -\Delta - \hbar\widetilde{\omega}(n+1) & v_\parallel\sqrt{2e\hbar B n} & 0 \\ 0 & v_\parallel\sqrt{2e\hbar B n} & \Delta + \hbar\widetilde{\omega}(n-1) & 0 \\ v_\parallel\sqrt{2e\hbar B n} & 0 & 0 & \Delta + \hbar\widetilde{\omega}(n+1) \end{pmatrix} \tag{2}$$

where $n$ is the Landau level index and $\widetilde{\omega} = eB/\widetilde{m}$. As for the effective mass corrections, the remote bands induce a small correction to the g-factor that accounts for $\hbar\widetilde{\omega}/2$ in this system [51,55]. The Hamiltonian (2) leads to Dirac-like Landau levels, which energies are given by:

$$E_n = \pm\hbar\widetilde{\omega} \pm \sqrt{(\Delta + \hbar\widetilde{\omega}n)^2 + 2e\hbar\left(v_\parallel^2 + \dfrac{\Delta}{\widetilde{m}}\right)Bn} \tag{3}$$

These energies were computed at $k_z = 0$ where the joint density of state of the interband magneto-optical transitions is maximal.

For the oblique valleys, the main ellipsoidal axes are not aligned with the magnetic field. It is therefore convenient to introduce a new coordinate system $(X,Y,Z)$ with the $Z$ axis parallel to $\bm{B}$, thus parallel to [111]. $(x,y,z)$ and $(X,Y,Z)$ are connected by a rotation combined with a scale change according to relative sizes of $v_\parallel$ and $v_z$. Using the Hamiltonian transformation as described in Refs. [41,42], the Landau levels of the oblique valleys are given by Eq. (3) in which $v_\parallel$ is replaced by [55,56]:

$$v_\| \left( \cos^2\theta + \frac{1}{K^2}\sin^2\theta \right)^{1/4}$$

The Landau levels are thus described by Eq. (3) for both types of valleys using two different electron velocities, which account for the anisotropy effect in the bulk sample.

### B. Quantum well Landau levels

Confinement effects in the quantum wells are calculated by adding a square potential $V(Z)$ into Eq. (1), where $Z//[111]$ is the heterostructure growth axis. The confinement potential $V(Z)$, whose value depends on the Eu content $y$, is electron-hole symmetric for PbSnSe/PbEuSe MQWs [36,44], as shown schematically in Fig. 1(b) where the variation of the band edges is drawn versus the growth direction. The potential $V(Z)$ makes [111] a preferential direction that will clearly differentiate the properties of the two types of valleys: longitudinal and oblique.

Let us first compute the energy and wavefunction of the $j^{th}$ confined state at $k_\| = 0$. To this end, we follow the procedure carried out in Ref. [44], which consists in solving the following equation and applying the probability current continuity conditions at each interface [57–59].

$$\begin{pmatrix} -\Delta + V(Z) - E^{(j)} & i\hbar \xi v_z \dfrac{d}{dZ} \\ -i\hbar \xi v_z \dfrac{d}{dZ} & \Delta + V(Z) - E^{(j)} \end{pmatrix} \begin{pmatrix} F_+^{(j)} \\ \xi F_-^{(j)} \end{pmatrix} = 0$$

where $\xi = \pm 1$ represents the two spin components and $F_\pm^{(j)}$ are the components on $L_6^\pm$ of the envelope function. Solving this equation leads to two confined states TIS and TIS' (plotted in blue in Fig. 1(b,c)), directly emerging from the topological character of the well material, as well as the usual confined states (plotted in orange) which emerge from the quantification of $L_6^\pm$. While TIS and TIS' wave functions are peaked at the interfaces, as shown in Fig. 1(c), the higher energy confined states (E2, H2, …) remain mainly localized in the heart of the QW, thus, representing "massive" QW states. If we consider the parity of the $F_+$ component of the $j^{th}$ confined state envelope wavefunction, TIS and E3 are odd, while E2 is even, and so on for the following confined states. As $F_-$ is proportional to $dF_+/dZ$, the parity of $F_-$ follows the opposite interplay.

For the calculation of the in-plane motions of the longitudinal valley, no Hamiltonian transformation is needed. We treat the $k_\|$-terms (or $B$-terms if we consider an applied magnetic field) in perturbation. It requires to calculate the matrix elements $\langle F_\pm^{(i)} | \delta W | F_\pm^{(j)} \rangle$ where $1 \leq i,j \leq N$ with $N$ the number of confined states obtained by solving the previous equation for $k_\| = 0$, and $\delta W$ is given by:

$$\delta W = \begin{pmatrix} -\dfrac{p_\|^2}{2\widetilde{m}}\mathbb{1} & v_\| \boldsymbol{p}_\| \cdot \boldsymbol{\sigma}_\| \\ v_\| \boldsymbol{p}_\| \cdot \boldsymbol{\sigma}_\| & \dfrac{p_\|^2}{2\widetilde{m}}\mathbb{1} \end{pmatrix}$$

For an applied magnetic field $\boldsymbol{B}//[111]$, the numerical solutions yield the Landau levels of the longitudinal valley. Those originating from the conduction band states are shown in Fig. 3(a) for MQW-39. Symmetric (i.e. related to energies of opposite sign) Landau levels are obtained for the confined hole states, as $V(Z)$ is electron-hole symmetric. Note that no interactions are obtained between Landau levels.

For the oblique valleys, an axis rotation is needed to account for their tilt with respect to the [111] orientation of the samples [56], as described in detail in Appendix A. As a result, the following Hamiltonian is obtained:

$$\begin{pmatrix} -[\Delta - V(Z)]\mathbb{1} & v_z P_Z \Sigma_Z \\ v_z P_Z \Sigma_Z & [\Delta + V(Z)]\mathbb{1} \end{pmatrix} + \begin{pmatrix} -\frac{P_\parallel^2}{2\widetilde{m}}\mathbb{1} & v_\parallel \boldsymbol{P}_\parallel \cdot \boldsymbol{\Sigma}_\parallel \\ v_\parallel \boldsymbol{P}_\parallel \cdot \boldsymbol{\Sigma}_\parallel & \frac{P_\parallel^2}{2\widetilde{m}}\mathbb{1} \end{pmatrix} + \sin\theta \begin{pmatrix} -\frac{1}{2\widetilde{m}}h(X,Z) & (v_\parallel - v_z)\mathcal{H}(X,Z) \\ (v_\parallel - v_z)\mathcal{H}(X,Z) & \frac{1}{2\widetilde{m}}h(X,Z) \end{pmatrix} \quad (4)$$

Here, $\boldsymbol{P}$ and $\boldsymbol{\Sigma}$ are the momentum and Pauli operators in the new coordinate system $(X, Y, Z)$. The two first terms of Eq. (4) are identical to those of the longitudinal valley. The third term includes the anisotropy effects with "$(v_\parallel - v_z)$" as well as the tilt effect with "$\sin\theta$". It is proportional to $\sin\theta$, and thus cancels for the longitudinal valley. Its diagonal elements write as:

$$\pm \frac{1}{2\widetilde{m}} h(X,Z) = \pm \frac{1}{2\widetilde{m}} [\sin\theta(P_Z^2 - P_X^2) - 2\cos\theta P_Z P_X]\mathbb{1} \quad (5)$$

These diagonal terms come from the anisotropic far-band correction terms that are $\widetilde{m}$ in the $XY$ plane and $\widetilde{\mu} \sim \infty$ along $Z$. Similarly, the off-diagonal elements are proportional to $(v_\parallel - v_z) = v_z(K-1)$ and account for the anisotropy effect of the valence and conduction bands. They are given by:

$$(v_\parallel - v_z)\mathcal{H}(X,Z) = (v_\parallel - v_z)[(\sin\theta P_Z - \cos\theta P_X)\Sigma_Z - (\sin\theta P_X + \cos\theta P_Z)\Sigma_X] \quad (6)$$

Solving Eq. (4) in the framework of the perturbation theory, which is detailed in Appendix B, yields the subband dispersions of the oblique valleys as well as the corresponding Landau levels. The intricate behavior of these Landau levels in which a large number of anti-crossings appears is revealed in Fig. 3(b). We focus on the first two anti-crossings zoomed-in in Fig. 3(c). They occur between Landau levels originating from TIS and E2 or E3 confined states. More generally, all these anti-crossings stem from two distinct kinds of couplings $W_1$ and $W_2$. The first coupling $W_1$ between the $i^{th}$ and the $j^{th}$ confined states is given by:

$$|W_1(nB)| = \sqrt{\frac{\hbar e B n}{2}} \frac{\sin 2\theta}{2} \left| \int_{-\infty}^{+\infty} \left[ \frac{\hbar}{\widetilde{m}} \left( F_+^{(i)} \frac{dF_+^{(j)}}{dZ} + F_-^{(i)} \frac{dF_-^{(j)}}{dZ} \right) - (v_\parallel - v_z) \left( F_+^{(i)} F_-^{(j)} - F_+^{(j)} F_-^{(i)} \right) \right] dZ \right| \quad (7)$$

The parities of $F_\pm^{(i)}$ and $F_\pm^{(j)}$ impose $j = i \pm 1$ so that this coupling involves two successive confined states. The anti-crossings take place between two Landau levels of successive subbands with indices $n$ and $n \pm 1$ and identical spins (see Appendix B). As an example, we highlight the anti-crossings that occur for magnetic field around 6 T and an energy about 40 meV between subbands TIS and E2 on Fig. 3(b,c).

The second potential $W_2$ is responsible for an interaction between TIS and E3 due to the parities of $F_\pm^{(i)}$ and $F_\pm^{(j)}$. It involves Landau levels with indexes $n$ and $n \pm 1$ and opposite spins. Such anti-crossings correspond for instance to the one computed around 14 T and 60 meV in Fig. 3(b,c). This second coupling $W_2$ is given by:

$$|W_2| = (v_\parallel - v_z)\hbar \frac{\sin 2\theta}{2} \left| \int_{-\infty}^{+\infty} \left[ F_+^{(i)} \frac{dF_-^{(j)}}{dZ} + F_-^{(i)} \frac{dF_+^{(j)}}{dZ} \right] dZ \right| \quad (8)$$

We highlight that $W_1$ and $W_2$ are proportional to the tilt $\theta$ of the valley axis with respect to the growth direction and to the valley anisotropy $(v_\parallel - v_z)$. Note also that the confinement, whether it occurs in the longitudinal or in the oblique valleys, leads to different energy quantization of the subbands, as shown by the extrapolations to $B = 0$ of the distinct Landau level series of Fig. 3(a,b). This "valley splitting" is due to the valley-dependent effective mass along the confinement direction [36,60]. For instance, the valley splitting for the confined state E4 is calculated as 3 meV.

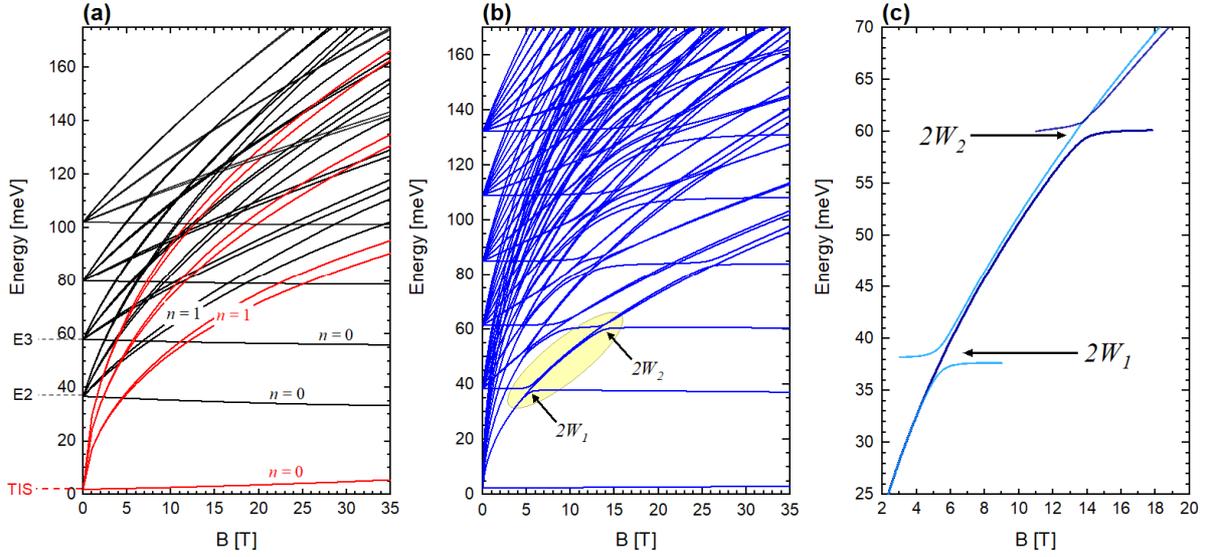

**Figure 3. (a)** Calculated Landau level energies of confined states at the $\bar{\Gamma}$-point for MQW-39 for parameters given in Table II. The Landau level from TIS (massive states) are drawn in red (black). Some of the Landau level indices are indicated. **(b)** Landau level energies of confined states at the $\bar{M}$-point for MQW-39. **(c)** Expanded view of the region encompassed by the yellow contour shown in (b). Dark and bright blue lines refer the different spin states. The anti-crossings due to the two coupling potentials $W_1$ and $W_2$ are highlighted.

As a remark, it is found that the slope of the $n = 0$ Landau levels depend on the valleys and are different for TIS and massive states. This slope scales as $\hbar e/\tilde{m}$ at the $\bar{\Gamma}$-point and as $\hbar e\cos^2\theta/\tilde{m}$ for oblique valleys. Therefore, as seen in Fig. 3(b), the slope of the $n = 0$ Landau levels are drastically reduced at the $\bar{M}$-points because of the anisotropic far-bands contribution. Furthermore, this slope is negative for massive states and positive for TIS. This effect accounts for the band symmetry of the 2D confined states. The massive states stem from the QW material bands, which are inverted, while TIS and TIS' emerge in a trivial ordering in IV-VI. As the remote bands symmetry ordering remain identical, their effects on $L_6^\pm$ are constant and tend to push up (down) in magnetic field a $n = 0$ Landau level that has a $L_6^-$ ($L_6^+$) symmetry. Thus, sign of the $n = 0$ Landau level slope is a direct indication of the confined state Bloch function symmetries. The far-bands contribution, which is sometimes called "inversion parameter", is therefore a good marker of band inversion in this system.

## IV. RESULTS AND DISCUSSION

The model derived for bulk $Pb_{1-x}Sn_xSe$ samples (see Sec. III.A) is used to interpret the magneto-optical fan chart presented in Fig. 2(a). Only one Dirac velocity is used to fit the data with Eq. (3), meaning that no anisotropy is measured. The absence of valley splitting in the magneto-optical data of the bulk sample implies that indeed $Pb_{0.75}Sn_{0.25}Se$ is a quasi-isotropic system as we pointed out previously [33]. In the following, we show, however, that in $Pb_{1-x}Sn_xSe/Pb_{1-y}Eu_ySe$ MQWs the barrier material causes a non-negligible anisotropy of the matrix elements.

The barrier material $Pb_{1-x}Eu_ySe$ is known to be anisotropic [36,37]. The electron velocities in $Pb_{1-y}Eu_ySe$ were determined as $v_\parallel \sim 6.5 \times 10^5$ m/s and $v_z = 3\sqrt{1.96 - 3.98\,y} \times 10^5$ m/s, so that $v_z \sim 3.75 \times 10^5$ m/s for MQW-39 ($K = v_\parallel/v_z = 1.73$) and $v_z \sim 4.00 \times 10^5$ m/s for MQW-25 ($K = $

1.63) [36]. The confined states in $Pb_{1-x}Sn_xSe/Pb_{1-y}Eu_ySe$ MQWs, whose probability densities in the barrier are non-negligible (calculated to be 16 and 10 % for MQW-25 and MQW-39 respectively), are sensitive to this anisotropy. In particular, TIS and TIS', whose probability densities are peaked at the interface (see Fig. 1(c)), are expected to be significantly influenced by the valley anisotropy of the barriers. The electron velocities of the quantum well states $v_\parallel$ and $v_z$ depict averaged velocities between those of the quantum well and barrier materials. They contain the effect of the barrier anisotropy and are treated as fitting parameters.

We use our theoretical model with $v_\parallel \neq v_z$ to interpret the magneto-optical data of MQW-39 shown in Fig. 2(b) and replotted in Fig. 4(a) where the calculated intersubband transitions between Landau levels from longitudinal ($\theta = 0°$) and oblique valleys ($\theta = 70.5°$) are shown by dashed and solid lines, respectively. Magneto-optical transitions occur between an electron-like and a hole-like confined states of opposite parities, thus, the allowed transitions involve for instance TIS' and TIS, H2 and E2, … Otherwise, the selection rules are similar to that obtained in the bulk case ($n \to n \pm 1$ and a spin flip) in the Faraday geometry. The absorptions shown by red dots are attributed to transitions between TIS' and TIS and those in black dots correspond to transitions involving excited states (H2, E2, E3, …) (see Fig. 3(a,b)). The observed magneto-optical absorptions mainly correspond to oblique valley transitions as they are three times more numerous than the longitudinal valley.

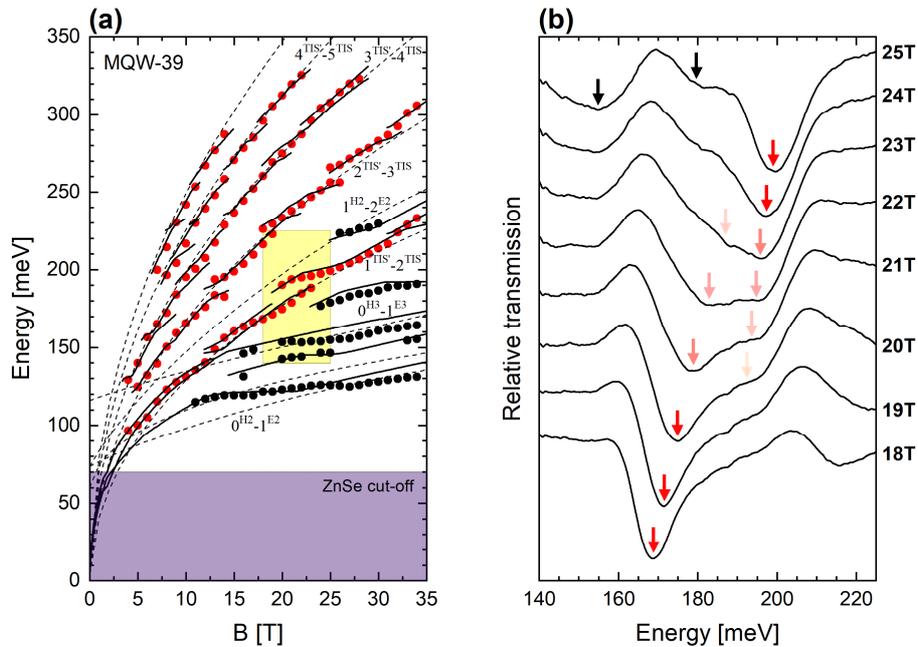

**Figure 4.** (a) Magneto-optical fan chart of MQW-39. Dots are experimental transmission minima and lines are calculated transitions between Landau levels. The red (black) color denotes absorptions that mainly involve TIS (massive states). The solid lines correspond to the oblique transitions and the dashed lines to the longitudinal valley transitions. The transitions labelling is only efficient for the longitudinal valley (see Fig. 3). (b) Magneto-optical spectra within the yellow rectangle in (a), highlighting the characteristic anti-crossings. The two absorptions are marked by arrows whose transparencies mimic absorption intensities.

The oblique valley transitions nicely fit the non-monotonic behavior of the Landau level transitions observed in the MQW sample, which is unequivocally due to the tilt and anisotropy of the oblique valleys that induce Landau level anti-crossings. The unusual dispersions of the absorptions (red and black dots) versus magnetic field are clearly well-described by taking into account the interactions between TIS and massive states, as elaborated in Sec. III. For instance, the second red absorption line, which involves TIS' and TIS, has an oscillatory behavior versus magnetic field. This is due to the interactions of the Landau levels of the TIS with the ones of E2, then E3, etc, as the magnetic field is increased. These interactions not only change the transition energies versus magnetic field but also yield a relaxation of the selection rules near the Landau level anti-crossings. To illustrate this effect, Figure 4(b) highlights one of these interactions occurring at around $B = 20$ T and photon energies around 180 meV. The two red arrows account for the swing of the intensity between two absorption lines, which is typical of Landau levels anti-crossings [52,61]. Note that it is difficult to model quantitatively the absorption intensities as a function of the magnetic field as the large number of involved Landau levels are strongly mixed between subbands, spins and their indices. Therefore, our model provides very accurate transition energies but it is rather difficult to provide a quantitative description of their intensities. These anti-crossings are clear experimental manifestations and proofs of the anisotropic and tilted valleys found in the investigated multivalley MQWs.

The parameters used to fit the magneto-optical absorptions in MQW-39 are listed in Table II. The slight discrepancy of the gap values between the well material and Pb$_{0.75}$Sn$_{0.25}$Se thick film can be explained by small Sn content variations or/and residual strain imposed by the barriers in the MQWs. The velocities are found to be slightly different for both types of valleys. Note that the anisotropy induced by the barriers has no reason to be similar for the two types of valleys [60,62]. Indeed, the penetration depth of the envelope functions depends on $v_z$, which is valley dependent [44].

**Table II.** Band parameters determined by the fit of the magneto-optical data in the three investigated samples.

| Parameter | Bulk | MQW-39 | MQW-25 |
|---|---|---|---|
| $2\Delta$ [meV] | -52.5 | -45 | -55 |
| $V$ [meV] | N.A. | 250 | 175 |
| $\widetilde{m}$ [$m_0$] | 0.35 | 0.35 | 0.35 |
| $v_\parallel^l$ [$10^5$ m/s] | 4.50 | 4.60 | 4.70 |
| $v_z^l$ [$10^5$ m/s] | 4.50 | 4.40 | 4.40 |
| $K_{\bar{\Gamma}}$ | N.A | 1.00 | 1.00 |
| $v_\parallel^o$ [$10^5$ m/s] | 4.50 | 4.90 | 4.95 |
| $v_z^o$ [$10^5$ m/s] | 4.50 | 3.90 | 4.15 |
| $K_{\bar{M}}$ | N.A | 1.23 | 1.17 |

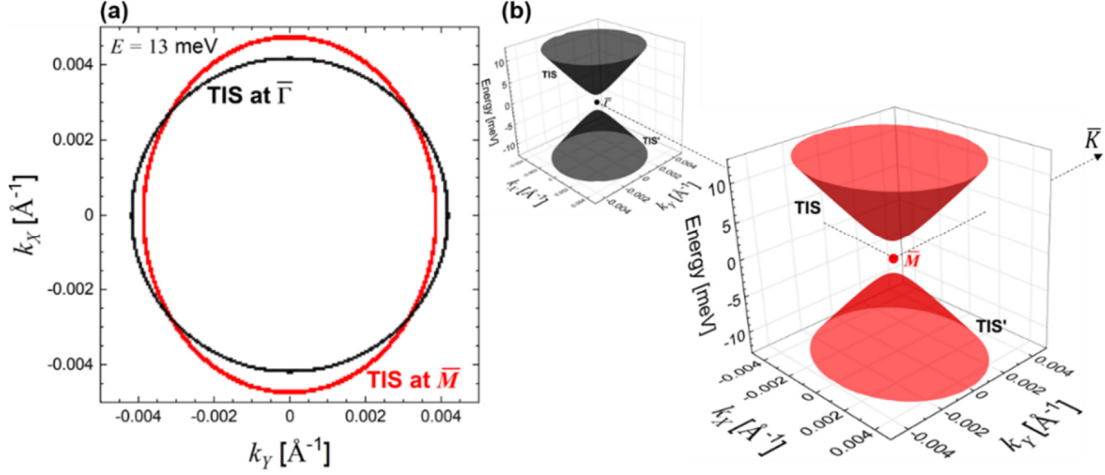

**Figure 5. (a)** Derived constant energy contours of TIS at $\bar{M}$ (in red) and $\bar{\Gamma}$ (in black) for arbitrarily chosen $E = 13$ meV and MQW-39. It is directly deduced from the experimental observation. **(b)** Map of the 2D Brillouin zone with the calculated dispersions of the MQW-39 TIS and TIS' near the $\bar{M}$ and $\bar{\Gamma}$ points.

Our analysis and the parameters determined in Table II lead to the accurate determination of the TIS anisotropy in the XY plane [12,24]. Because of the anisotropy and different tilts of the 3D ellipsoids (see Fig. 1(a)), their projection on the 2D BZ are not equivalent at the $\bar{M}$-points and at the $\bar{\Gamma}$-point, leading to anisotropic constant energy contours, as sketched in Fig. 5(a). Remarkably, this projection due to confinement allows for a different anisotropy factor at $\bar{\Gamma}$ and $\bar{M}$, which is not possible with bulk materials where the anisotropy factor $K$ is identical for oblique and longitudinal valleys. We thus define the two anisotropy factors $K_{\bar{M}}$ and $K_{\bar{\Gamma}}$ associated with the constant energy surfaces of the TIS located at $\bar{M}$ and $\bar{\Gamma}$. They are determined by the ratios between the in-plane velocities in the $Y$ and $X$ directions and are listed in Table II. The coordinates rotation detailed in Appendix A defines $\bar{M}$-$\bar{\Gamma}$ as the $X$ direction and $\bar{M}$-$\bar{K}$ as the $Y$ axis (see Fig. 1(a)). The expression for the TIS dispersions is given in Appendix C. For the longitudinal valley, we obtain an isotropic dispersion in the $XY$ plane, with the velocity $v_{\parallel}^{l}$. Therefore, one gets $K_{\bar{\Gamma}} = 1$ at the $\bar{\Gamma}$-point, which follows directly from the isotropy of the well and barrier materials in the layer plane (see Eq. (1)) for this geometry. However, the TIS dispersions near the $\bar{M}$-points are anisotropic with electron velocities $v_{\parallel}^{o}$ along $\bar{M}$-$\bar{K}$ and $\cos^2\theta v_{\parallel}^{o} + \sin^2\theta v_{z}^{o}$ in the $\bar{M}$-$\bar{\Gamma}$ directions (see Appendix C). The magneto-optical determination of $v_{\parallel}^{o}$ and $v_{z}^{o}$ by fitting the anti-crossing magnitudes allows the deduction of the topological state dispersion anisotropy, which is given by:

$$K_{\bar{M}} = \left[\cos^2\theta + v_{z}^{o}\sin^2\theta / v_{\parallel}^{o}\right]^{-1}$$

For the sample MQW-39, we find $K_{\bar{M}} = 1.23$, which yields the constant energy surface shown in red in Fig. 5(a) at 13 meV. For comparison, the isotropic constant energy surface of the longitudinal TIS is drawn in black. The TIS dispersions near the $\bar{M}$ and $\bar{\Gamma}$ points are shown in Fig. 5(b). This data demonstrates that magneto-spectroscopy is a powerful technique to determine accurately the anisotropy of the massive or massless Dirac cones (TIS and TIS') at the $\bar{M}$-points.

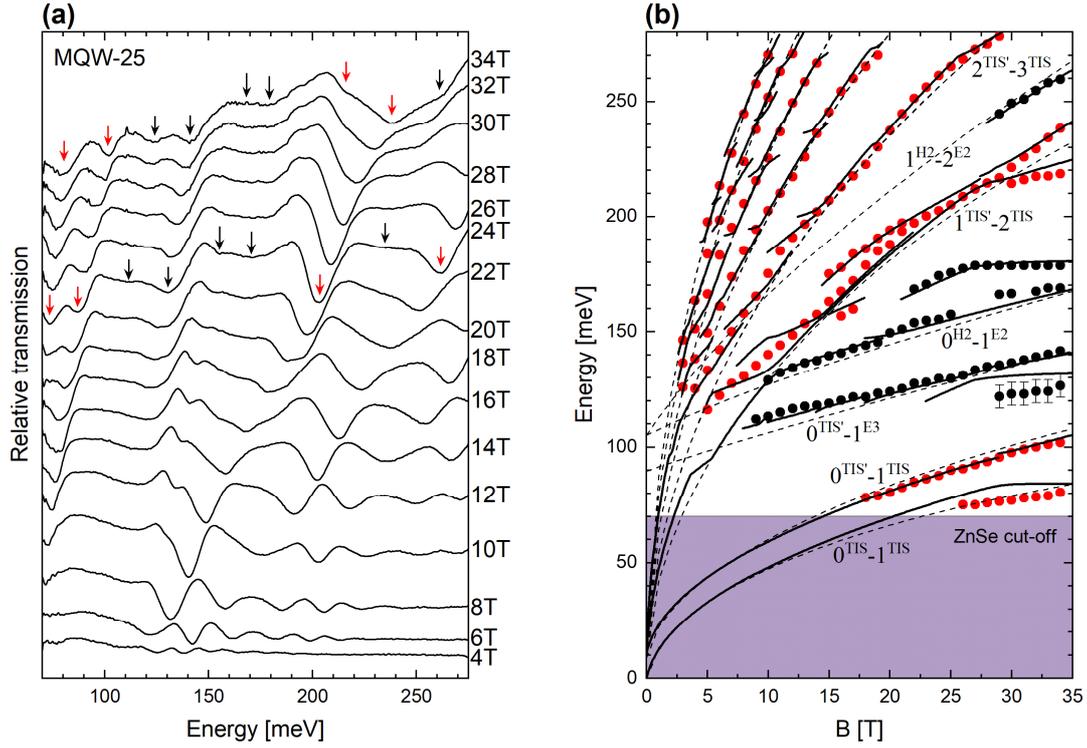

**Figure 6. (a)** Magneto-optical spectra of MQW-25. Some transmission minima are marked by colored arrows. **(b)** Magneto-optical fan chart of MQW-25. Dots represent transmission minima pointed out in (a). Those in red represent absorptions mainly involving TIS and TIS', while the black dots are mainly due to massive states. Dashed (solid) lines are the fit from the Landau level transitions emerging from the longitudinal valley (oblique valleys). The transition labels are given for the longitudinal valley.

The magneto-optical spectra up to 34 T obtained on the second MQW sample (MQW-25) are shown in Fig. 6(a) and the interpretation is summarized by Fig. 6(b). The solid lines are transitions between Landau levels of the oblique valleys and the dashed lines account for the longitudinal one. The fitting parameters are listed in Table II and are found to be similar as those of MQW-39. Again, our theoretical model remarkably accounts for the observed magneto-optical transitions, including several anti-crossings between TIS and massive state Landau levels emerging at the $\bar{M}$ points. For this sample MQW-25 the anisotropy factor for the TIS dispersion at the $\bar{M}$-point is deduced as $K_{\bar{M}} = 1.17$ and is found to be slightly lower than the one of MQW-39. These results confirm our analysis and fully demonstrate the interaction of TIS Landau levels with those of the massive states in the MQWs in the presence of valley tilt and anisotropy.

## V. CONCLUSION

We have demonstrated significant interactions between topological and massive states in multivalley TCI quantum well heterostructures. The anisotropy induced by the barrier normal insulator allows to differentiate between the electronic properties of the topological states emerging from the longitudinal and oblique valleys. The Landau levels associated to the oblique valleys show anti-crossings between topological interface states and massive states near the $\bar{M}$-points of the 2D BZ, which is not the case for Landau levels emerging from the longitudinal valley, near the $\bar{\Gamma}$-point. Moreover, our work allows us to accurately determine the anisotropy of the topological states of both kinds of valleys. In this way, we demonstrate an isotropic, i.e., circular dispersion for TIS near $\bar{\Gamma}$ and a tunable elliptical anisotropy of TIS near the $\bar{M}$-points induced by the strong valley anisotropy within the barrier. Our work establishes a link between the parent 3D ellipsoids of bulk materials and their projection when a quantum confinement is superimposed, and thus, opens the door to valley and anisotropy engineering of topological states. In particular, the effects demonstrated in this work (Landau levels avoided crossings and TIS anisotropy) should also be observed in $Pb_{1-x}Sn_xTe$-based heterostructures, where the anisotropy of the TIS should be greatly enhanced as $Pb_{1-x}Sn_xTe$ is much more anisotropic ($K \sim 3.2$) compared to $Pb_{1-x}Sn_xSe$ [53,61,63].

Our work thus paves the way to valley engineering in topological insulators and to future investigations showing valley effects that can spontaneously break crystalline symmetry, such as nematic valley ordering. The anisotropy of the TIS dispersions in $Pb_{1-x}Sn_xSe$ quantum wells is indeed predicted to promote a valley polarization that would drive the formation of quantum Hall ferroelectric states as theoretically anticipated for different material systems [15–17,64,65].

## ACKNOWLEDGEMENTS

The authors want to thank G. Bastard for enlightening discussions. This work is supported by a co-project between ANR N° ANR-19-CE30-022-01 for ENS and the Austrian Science Fund FWF, Project I-4493 for JKU. We acknowledge the support of LNCMI-CNRS, a member of the European Magnetic Field Laboratory (EMFL). M.O. acknowledges the support from ANR No-19-CE30-0032 project Collector.

## APPENDIX A: OBLIQUE VALLEY HAMILTONIAN IN HETEROSTRUCTURE

As pointed out in Sec. III.B, for an oblique valley with main axis $(x, y, z)$, the quantum operators have to be expressed in a new coordinate system $(X, Y, Z)$ where $Z//\mathbf{B}//[111]$. The relation between operators in both coordinate systems is encoded in a rotation matrix $\mathbb{M}$, which leads to $\mathbf{p} = \mathbb{M}\mathbf{P}$ and $\mathbf{\sigma} = \mathbb{M}\mathbf{\Sigma}$ for the momentum and spin operators respectively [56]. Using the invariance of a scalar product under rotation, the off-diagonal terms of the Hamiltonian (1) transform as:

$$v_\parallel \mathbf{p}_\parallel \mathbf{\sigma}_\parallel + v_z p_z \sigma_z \rightarrow v_\parallel \mathbf{P}_\parallel \mathbf{\Sigma}_\parallel + v_z P_Z \Sigma_Z + (v_\parallel - v_z)(P_Z \Sigma_Z - p_z \sigma_z)$$

with:

$$P_Z \Sigma_Z - p_z \sigma_z = m_{zx}[(m_{zx} P_Z - m_{zz} P_X)\Sigma_Z - (m_{zx} P_X + m_{zz} P_Z)\Sigma_X]$$

where $m_{\alpha\beta}$ denote the matrix elements of $\mathbb{M}$. The three oblique valleys are tilted by an angle $\theta$ with respect to $Z$ (see Fig. 1(a)) so that $m_{zz} = \cos\theta$. As $\mathbb{M}\mathbb{M}^{-1} = \mathbb{1}$, one gets $m_{zx}^2 + m_{zy}^2 + m_{zz}^2 = 1$. The isotropic dispersion in the $xy$ plane allows to arbitrarily choose $m_{zy} = 0$ (i.e. choosing a rotation with respect to $y//[1\bar{1}0]$), which leads to $m_{zx} = |\sin\theta|$. One ends up with $P_Z\Sigma_Z - p_z\sigma_z = \sin\theta \mathcal{H}(X, Z)$ as defined in Eq. (6). A similar transformation is done for the diagonal terms $(p_\parallel^2/2\widetilde{m})\mathbb{1}$ of (1):

$$\frac{p_\parallel^2}{2\widetilde{m}}\mathbb{1} \rightarrow \frac{P_\parallel^2}{2\widetilde{m}}\mathbb{1} + \frac{1}{2\widetilde{m}}(P_Z^2 - p_z^2)\mathbb{1}$$

where $p_z = m_{zx} P_X + m_{zz} P_Z$. The resulting Hamiltonian is given by (4) as $(P_Z^2 - p_z^2)\mathbb{1} = \sin\theta\, h(X, Z)$ (Eq. (5)).

## APPENDIX B: OBLIQUE VALLEY LANDAU LEVELS

The perturbative terms for the oblique valleys (second and third terms of Eq. (4)) are given by:

$$\delta W = \begin{pmatrix} -\dfrac{P_\parallel^2}{2\widetilde{m}}\mathbb{1} - \dfrac{\sin\theta}{2\widetilde{m}} h(X,Z) & v_\parallel \mathbf{P}_\parallel \cdot \mathbf{\Sigma}_\parallel + \sin\theta(v_\parallel - v_z)\mathcal{H}(X,Z) \\ v_\parallel \mathbf{P}_\parallel \cdot \mathbf{\Sigma}_\parallel + \sin\theta(v_\parallel - v_z)\mathcal{H}(X,Z) & \dfrac{P_\parallel^2}{2\widetilde{m}}\mathbb{1} + \dfrac{\sin\theta}{2\widetilde{m}} h(X,Z) \end{pmatrix}$$

With $h(X, Z)$ and $\mathcal{H}(X, Z)$ given by Eq. (5) and (6) respectively. Under a magnetic field along the $Z$ direction, we can write:

$$\begin{cases} P_Z = -i\hbar\, d/dZ \\ P_X = \sqrt{\dfrac{e\hbar B}{2}}(a + a^+) \\ P_X + iP_Y = \sqrt{2e\hbar B}\, a^+ \\ P_X - iP_Y = \sqrt{2e\hbar B}\, a \end{cases}$$

where $a$ and $a^+$ are the ladder operators. The perturbative Hamiltonian cannot be projected into a trivial basis of harmonic oscillator functions, as it was the case for the longitudinal valley. The envelope function needs to be expressed in its general form and, for the $j^{th}$ confined state, it is given by:

$$\psi^{(j)} = \sum_{n,i} \left[ \alpha_{n,i} \begin{pmatrix} F_+^{(j)} \\ 0 \\ F_-^{(j)} \\ 0 \end{pmatrix} + \beta_{n,i} \begin{pmatrix} 0 \\ F_+^{(j)} \\ 0 \\ -F_-^{(j)} \end{pmatrix} \right] |n\rangle$$

$|n\rangle$ is the harmonic oscillator functions with $n = 0,1,2,\ldots$ $\alpha$ and $\beta$ are the coefficients in front of the two spin components. The perturbative theory requires to calculate matrix elements $\langle \psi^{(i)} | \delta W | \psi^{(j)} \rangle$, where $1 \leq i,j \leq N$ with $N$ the number of confined states obtained at $k_\parallel = 0$. This leads to two equations on $\alpha_{n,j}$ and $\beta_{n,j}$ that are coupled with $\alpha_{n\pm1,j}, \beta_{n\pm1,j}, \alpha_{n\pm2,j} \ldots$ We solved these equations by taking $1 \leq j \leq N$ and $0 \leq n \leq 8$, which insures a convergence. For $n = 0$, only the equation on $\beta_{0,j}$ is considered, the equation on $\alpha_{0,j}$ being unphysical in this system. Indeed, the two $n = 0$ Landau levels are spin polarized [51,52]. This resolution leads to the oblique valley Landau levels shown in Fig. 3(b). Moreover, we can extract from these equations the two coupling potentials $W_1$ and $W_2$. They occur between $\beta_{n+1,j}$ and $\beta_{n,j\pm(2q+1)}$ (or between $\alpha_{n+1,j}$ and $\alpha_{n,j\pm(2q+1)}$) for $W_1$ and between $\beta_{n,j}$ and $\alpha_{n+1,j\pm(2q+2)}$ for $W_2$ ($q \in \mathbb{N}$).

### APPENDIX C: ANISOTROPY OF THE TOPOLOGICAL INTERFACE STATES

The theoretical model detailed in Sec. III.B gives the perturbative matrix for the in-plane dispersion calculations. The resolution is similar to the one in Appendix B.

As an illustration, in the subspace TIS and TIS' (twofold degenerated due to the spin), the perturbation theory leads to the diagonalization of the following matrix: (for the sake of simplicity, small terms like remote band effects are neglected)

$$\begin{pmatrix} -\delta \Sigma_Z & -i[\hbar v_\parallel k_- - \hbar \sin^2\theta (v_\parallel - v_z) k_X] \Sigma_X \\ i[\hbar v_\parallel k_+ - \hbar \sin^2\theta (v_\parallel - v_z) k_X] \Sigma_X & -\delta \Sigma_Z \end{pmatrix}$$

Here, $\pm \delta$ are the energies of TIS' and TIS calculated at $k_\parallel = 0$; and $k_\pm = k_X \pm i k_Y$. In a good approximation, the TIS dispersions are thus twofold degenerated with:

$$E(\theta) = \pm \sqrt{\delta^2 + \hbar^2 v_\parallel^2 k_Y^2 + \hbar^2 \left( v_\parallel - \sin^2\theta (v_\parallel - v_z) \right)^2 k_X^2}$$

For the longitudinal valley we have $\theta = 0°$ so that the intersubband matrix element of momentum between TIS and TIS' is $v_\parallel$. At the $\overline{M}$-point, where $\theta = 70.5°$, one immediately obtains the in-plane velocities and the anisotropy factor $K_{\overline{M}}$ mentioned in Sec IV.

As a final remark, when a magnetic field along $Z$ is applied, the Hamiltonian above gives, in a good approximation, the Landau levels of TIS' and TIS for the two types of valley:

$$E_n(\theta) = \pm \sqrt{\delta^2 + 2e\hbar Bn \left[ \frac{v_\parallel + v_z}{2} + \frac{\cos^2\theta}{2} (v_\parallel - v_z) \right]^2}$$

Therefore, the in-plane ($XY$ plane) motions of the electrons under magnetic field in both valleys are characterized by the following velocity:

$$\left[ \frac{v_\parallel + v_z}{2} + \frac{\cos^2\theta}{2} (v_\parallel - v_z) \right]$$